\documentclass[10pt]{article}
\usepackage{amsfonts}
\usepackage{amsmath}
\usepackage{amssymb}
\usepackage{epsfig}
\usepackage{graphicx,color}
\usepackage{mathptmx}
\begin{document}
\title{Testing the Non-universal $Z^\prime$ Model
in $B_s \to \phi \pi^0$ Decay}
\author{~~Juan Hua$^{1}$\footnote{Email: juanhua@126.com},
~~C.~S. Kim$^{2}$\footnote{Email: cskim@yonsei.ac.kr}~,~~ Ying
Li$^{1,2}$\footnote{Email: liying@ytu.edu.cn} \\
 \small {\it  1. Department
of Physics, Yantai University, Yantai 264-005, China } \\
\small{\it 2.Department of Physics and IPAP, Yonsei University, Seoul
120-479, Korea }\\
} \maketitle
\begin{abstract}
\noindent The branching ratio and direct CP asymmetry of the decay
mode $B_s \to \phi \pi^0$ have been calculated within the QCD
factorization approach in both the standard model (SM) and the
non-universal $Z^\prime$ model. In the standard model, the CP
averaged branching ratio is about $1.3\times 10^{-7}$. Considering
the effect of $Z^\prime$ boson, we found the branching ratio can be
enlarged
three times or decreased to one third 
within the allowed parameter spaces. Furthermore, the direct CP asymmetry
could reach $55\%$ with a light $Z^\prime$ boson and suitable CKM
phase, compared to $25\%$ predicted in the SM. The enhancement
of both branching ratio and CP asymmetry cannot be realized at the
same parameter spaces, thus, if this decay mode is measured in
the upcoming LHC-b experiment and/or Super B-factories, the peculiar deviation from the SM may provide a
signal of the non-universal $Z^\prime$ model, which can be used to constrain the mass of
$Z^\prime$ boson in turn.
\end{abstract}

\newpage

Although most of the experimental data are consistent with the standard
model (SM) predictions, it is believed that the SM is just an
effective theory of a more fundamental one yet to be discovered. One
way of searching for new physics beyond the SM is by studying the
rare $B$ decay modes, which are induced by flavor changing neutral
current (FCNC) transitions, since such rare decays arise only from the
loop level within the SM. Over the years, many studies have been made to
predict the branching ratios and CP asymmetries of $B$ decays in the
SM and in new physics (NP) models, such as supersymmetry and etc.
Although the presence of NP in the $b$ sector is not yet
firmly established, there exist several signals which will be
verified in the forthcoming LHC-b experiment and super-B factories.
Therefore, it is interesting to explore as many rare decays as possible
to find an indication of NP.

Additional $\mathrm{U}(1)^\prime$ gauge symmetries and associated $Z^\prime$
gauge bosons \cite{Langacker:2000ju} could appear in several well motivated extensions of the SM.
Searching for an extra $Z^\prime$ boson is an important mission in the experimental
programs of Tevatron and LHC. One of the simple extensions beyond the SM is the
family non-universal $Z^\prime$ model, which could be
naturally derived in certain string constructions
\cite{Buchalla:1995dp}, E6 models \cite{Nardi:1992nq} and so on. It
is interesting to note that the non-universal $Z^\prime$ couplings
could lead to FCNC in the tree level as well as introduce new weak phases, which are essential
in inducing the CP asymmetries. The
effects of $Z^\prime$ in $B$ sector have been investigated in a
number of papers, such as Refs. \cite{Barger:2009hn,Cheung:2006tm}.
The recent review about $Z^\prime$ in detail is referred to Ref.
\cite{Langacker:2008yv}.

In this work, we will address the effect of the $Z^\prime$ in the
rare decay mode $B_s \to \phi \pi^0$. It is expected to have a small
branching ratio in the SM because it is an electro-weak penguin dominated
process and  mediated by  $b\to s q\bar
q$.  In dealing with the two body charmless non-leptonic $B$ decays,
many approaches have been proposed, such as the naive factorization,
the QCD factorization (QCDF) approach
\cite{Beneke:1999br,Beneke:2003zv}, the perturbative QCD (PQCD) approach
and the soft collinear effective theory (SCET). In previous studies, the
branching ratio is shown to be about $10^{-7}$ in the SM, both in the QCDF approach
\cite{Beneke:2003zv} and in the PQCD approach \cite{PQCD}.  For completeness,  we would first calculate the mode
within the SM, before discussing the effect of the new physics. Since there is no annihilation
contribution in this decay, we will adopt the QCDF approach.

We start from the relevant effective Hamiltonian given by:
\begin{eqnarray}\label{hamiton}
 {\cal H}_{eff}={G_F\over \sqrt 2}\Big[V_{ub}V^*_{us}\Big(
 C_1O_1^p+C_2O_2^p\Big)-V_{tb}V^*_{ts}\sum\limits_{i=3}^{10}
 C_iO_i\Big].
\end{eqnarray}
The explicit form of the operators $O_i$  and the corresponding
Wilson coefficients $C_i$ at the scale of  $\mu=m_b$ can be found in
Ref. \cite{Buras}. $V_{u(t)b}$, $V_{u(t)s}$ are the
Cabibbo-Kabayashi-Maskawa (CKM) matrix elements.

In the QCDF approach, the contribution of the non-perturbative
sector is dominated by the form factors of $B_s\to \phi $ transition
and the non-factorizable impact in  the hadronic matrix elements is
controlled by hard gluon exchange.  The hadronic matrix elements of
the decay can be written as
\begin{eqnarray}
 \langle \phi \pi|O_i|B\rangle &=& \sum_{j}F_{j}^{B_s\rightarrow
\phi }\int_{0}^{1}dx T_{ij}^{I}(x)\Phi_{\pi}(x) \nonumber\\
 &+&\int_{0}^{1}d\xi\int_{0}^{1}dx\int_{0}^{1}dy
T_{i}^{II}(\xi,x,y)\Phi_{B}(\xi)\Phi_{\phi}(x)\Phi_{\pi}(y).
\end{eqnarray}
Here  $T_{ij}^{I}$ and $T_{i}^{II}$ denote the perturbative
short-distance interactions and can be calculated perturbatively.
$\Phi_{X}(x)~(X=B_s,\pi,\phi)$ are the universal and
non-perturbative light-cone distribution amplitudes, which can be
estimated by the light cone QCD sum rules.
Following the standard procedure of QCD factorization approach, we
can write the decay amplitude as
\begin{eqnarray} \label{eq:af} {\cal A}(\overline{B_s^0}\to
\phi\pi^0)=\frac{G_F}{\sqrt{2}}
\sum_{p=u,c}\sum_{i}V_{pb}V_{ps}^{*}a_i^p(\mu)\langle
\phi\pi^0|O_i|B\rangle,
\end{eqnarray}
where $\langle M_{1}M_{2}|O_i|B\rangle _{F} $ is the factorizable
matrix element, which can be factorized into a form factor times a
decay constant, and the coefficients
$a_i$ ($i=1$ to $10$) can be found in Refs. \cite{Beneke:1999br,Beneke:2003zv}.
Note that in dealing with the hard-scattering spectator interactions in the QCDF,
there is an infrared endpoint singularity, which can only be
estimated in a model-dependent way  with a large uncertainty. In
Refs. \cite{Beneke:1999br,Beneke:2003zv}, this contribution is
parameterized by  one complex quantity $X_H$,
\begin{eqnarray}
X_{H} = \left ( 1 + \rho_{H} e^{i\phi_{H}}\right )
\ln\frac{m_B}{\Lambda_h}~,
\end{eqnarray}
where $\Lambda_h =0.5$ GeV, $\phi_{H}$ is a free strong phase in the
range $[-180^\circ,180^\circ]$, and $\rho_{H}$ is a real parameter
varying within $[0,1]$.

Finally the decay amplitude can be given as
\begin{multline} \label{eq:af1}
{\cal A}(\overline{B_s^0}(p_B)\to \phi(\epsilon,p_1) \pi^0(p_2) )=-
i\frac{G_F}{2} 2 m_\phi f_\pi(\epsilon^*\cdot p_B) A_0^{B_s\to
\phi}(0)\\ \times\Big[V_{ub}V_{us}^*
(a_2[u]+a_3[u]-a_3[d]-a_5[u]+a_5[d]-a_7[u]-\frac{1}{2}a_7[d]+a_9[u]+\frac{1}{2}a_9[d])\\
+V_{cb}V_{cs}^*(a_3[u]-a_3[d]-a_5[u]+a_5[d]-a_7[u]-\frac{1}{2}a_7[d]+a_9[u]+\frac{1}{2}a_9[d])\Big],
\end{multline}
where the symbols $u$ and $d$ in square brackets indicate the
component of the meson $\pi^0$. In the SM, $a_i[u]=a_i[d]=a_i$, therefore,
we get the  simplified formula for the decay amplitude:
\begin{eqnarray} \label{eq:af2}
{\cal A}(\overline{B_s^0}\to \phi\pi^0)=- i\frac{G_F}{2}m_B^2 f_\pi
A_0^{B_s\to \phi}(0) \times\Big[V_{ub}V_{us}^*
(a_2-\frac{3}{2}a_7+\frac{3}{2}a_9)+V_{cb}V_{cs}^*(-\frac{3}{2}a_7+\frac{3}{2}a_9)\Big],
\end{eqnarray}
after utilizing $2 m_\phi (\epsilon^*\cdot p_B)=m_B^2$. The branching
ratio takes the form
\begin{eqnarray}
{\cal B}(\overline{B_s^0}\to \phi\pi^0)=\tau_B\frac{|P_c|}{8\pi
M_B^2}|{\cal A}(\overline{B_s^0}\to \phi\pi^0)|^2\, ,
\end{eqnarray}
where $\tau_B$ is the $B_s$ meson lifetimes, and $ |P_c|$ is the
absolute value of two final-state hadrons' momentum in the $B_s$
rest frame. We can also define the direct CP asymmetry as:
\begin{eqnarray}
A_{CP}=\frac{|{\cal A}(\overline{B_s^0}\to \phi\pi^0)|^2-|{\cal
A}(B_s^0\to \phi\pi^0)|^2}{|{\cal A}(\overline{B_s^0}\to
\phi\pi^0)|^2+|{\cal A}(B_s^0\to \phi\pi^0)|^2}.
\end{eqnarray}
Note that in the naive factorization there is no CP asymmetry because of
none existence of any strong phase, which is a key factor in producing a direct CP
asymmetry.

\begin{table}[b]
\begin{center}
\caption{Summary of input parameters \cite{Beneke:2003zv}}
\label{parameter}
\begin{tabular}{c}\hline\hline
 $\begin{array}{ccccccccc}
 \lambda & A &\bar{\rho}& \bar{\eta}&\Lambda_{\overline{\mathrm{MS}}}^{(f=4)}
 &\tau_{B_s^0}&\lambda_B&\alpha_e&\alpha_s\\
 0.225& 0.818&0.141& 0.348& 250
 \mbox{MeV}&1.46\mbox{ps}&0.35&1/132 &0.214
 \end{array}$\\
 \hline
 $\begin{array}{ccccccccc}
f_{B_s}      &  m_{B_s}         & f_{\pi} & f_{\phi} &f_{\phi}^\perp &m_\phi& \gamma&A_0^{B_s\to \phi} & \\
236\mbox{MeV}& 5.36\mbox{GeV}&  131\mbox{MeV}&221\mbox{MeV} &175\mbox{MeV}&1.01\mbox{GeV} &70^\circ &0.34  &
 \end{array}$ \\
   \hline\hline
\end{tabular}
\end{center}
\end{table}
For the numerical calculation, with the input parameters listed in
Table. \ref{parameter}, the averaged branching ratio and direct CP
asymmetry of decay $B_s\to \phi\pi^0$ obtained in the SM are
\begin{eqnarray}
{\cal B}(B_s\to \phi\pi^0)&=&1.3\times 10^{-7}\, ,\nonumber\\
A_{CP}(B_s\to \phi\pi^0)&=&25\%\, ,
\end{eqnarray}
which have not yet been measured in the Tevatron experiments. However,  the
order of magnitudes should be measured easily in the LHC-b experiment and/or Super
B-factories in future. Because we used the updated parameters, the
branching ratio is slightly larger than that predicted in
Ref. \cite{Beneke:2003zv}, and the CP asymmetry agrees with each
other. The results also agree with the predictions from the PQCD
\cite{PQCD} as well. Here we will not tend to discuss the
uncertainties in our calculation, since this part has been
presented explicitly in  \cite{Beneke:2003zv}.

Now we  turn to  the effects due to an extra
$U(1)^\prime$ gauge boson $Z^\prime$. We start from the interactions
with the new $Z^\prime$ gauge particle ignoring the mixing between
$Z_0$ and $Z^\prime$. Following the convention in
Ref. \cite{Langacker:2000ju}, we write the couplings of the
$Z^\prime$-boson to fermions as
\begin{eqnarray}
J_{Z^{\prime}}^{\mu}=g^{\prime}\sum_i \bar\psi_i
\gamma^{\mu}[\epsilon_i^{\psi_L}P_L+\epsilon_i^{\psi_R}P_R]\psi_i,
\label{eq:JZprime}
\end{eqnarray}
where $i$ is the family index and $\psi$ labels the fermions  and
$P_{L,R}=(1\mp\gamma_5)/2$. According to  certain string constructions
\cite{chaudhuri} or GUT models \cite{gut}, it is possible to have
family non-universal $Z^{\prime}$ couplings. That is, even though
$\epsilon_i^{L,R}$ are diagonal, the couplings are not family
universal. After rotating to the physical basis, FCNC's generally
appear at tree level in both left handed and right handed sectors,
explicitly, as
\begin{eqnarray}
B^{\psi_L}=V_{\psi_L}\epsilon^{\psi_L}V_{\psi_L}^{\dagger},\;\;\;\;\;
B^{\psi_R}=V_{\psi_R}\epsilon^{\psi_R}V_{\psi_R}^{\dagger}.
\end{eqnarray}
For simplicity,  we assume that the right-handed couplings are
flavor-diagonal and neglect $B_{sb}^R$, thus the $Z^{\prime}$ part
of the effective Hamiltonian for $b\to s\bar{q}q\,(q=u,d)$
transitions has the form as:
\begin{equation}\label{heffz1}
 {\cal H}_{eff}^{\rm
 Z^{\prime}}=\frac{2G_F}{\sqrt{2}}\big(\frac{g^{\prime}M_Z}
 {g_1M_{Z^{\prime}}}\big)^2
 \,B_{sb}^L(\bar{s}b)_{V-A}\sum_{q}\big(B_{qq}^L (\bar{q}q)_{V-A}
 +B_{qq}^R(\bar{q}q)_{V+A}\big)+h.c.\,,
\end{equation}
where $g_1=e/(\sin{\theta_W}\cos{\theta_W})$ and $M_{Z^{\prime}}$ is
the new gauge boson mass. Compared with the operators existed in the
SM,  Eq.~(\ref{heffz1}) can be modified as
\begin{equation}
 {\cal H}_{eff}^{\rm
 Z^{\prime}}=-\frac{G_F}{\sqrt{2}}V_{tb}V_{ts}^{\ast}\sum_{q}
 (\Delta C_3 O_3^q +\Delta C_5 O_5^q+\Delta C_7 O_7^q+\Delta C_9
  O_9^q)+h.c.\,,
\end{equation}
where $O_i^q(i=3,5,7,9)$ are the effective operators in the SM, and
$\Delta C_i$ the modifications to the corresponding SM Wilson
coefficients caused by $Z^{\prime}$ boson, which are expressed as
\begin{eqnarray}
 \Delta C_{3}&=&-\frac{2}{3V_{tb}V_{ts}^{\ast}}\,\big(\frac{g^{\prime}M_Z}
 {g_1M_{Z^{\prime}}}\big)^2\,B_{sb}^L\,(B_{uu}^{L}+2B_{dd}^{L})\,,\nonumber\\
 \Delta C_{5}&=&-\frac{2}{3V_{tb}V_{ts}^{\ast}}\,\big(\frac{g^{\prime}M_Z}
 {g_1M_{Z^{\prime}}}\big)^2\,B_{sb}^L\,(B_{uu}^{R}+2B_{dd}^{R})\,,\nonumber\\
 \Delta C_{7}&=&-\frac{4}{3V_{tb}V_{ts}^{\ast}}\,\big(\frac{g^{\prime}M_Z}
 {g_1M_{Z^{\prime}}}\big)^2\,B_{sb}^L\,(B_{uu}^{R}-B_{dd}^{R})\,,\nonumber\\
 \Delta C_{9}&=&-\frac{4}{3V_{tb}V_{ts}^{\ast}}\,\big(\frac{g^{\prime}M_Z}
 {g_1M_{Z^{\prime}}}\big)^2\,B_{sb}^L\,(B_{uu}^{L}-B_{dd}^{L})\,,
 \label{NPWilson}
\end{eqnarray}
in terms of the model parameters at the $M_W$ scale. While we can
have $Z^\prime$ contributions to the QCD penguins as well as the EW
penguins, in view of the results evaluated by Buras et. al
\cite{ajburas}, we set $B_{uu}^{L,R}=-2 B_{dd}^{L,R}$, so that new
physics is manifest in the EW penguins.
 Without loss of generality, we always assume that the diagonal
elements of the effective coupling matrices $B_{qq}^{L,R}$ are real
due to the hermiticity of the effective Hamiltonian. However, there still is a
new weak phase $\phi$ in the off-diagonal one of $B_{sb}^{L}$. The
resulting $Z^\prime$ contributions to the Wilson coefficients are:
\begin{eqnarray}
& &\Delta C_{3,5}\simeq 0, \nonumber\\
& &\Delta
C_{9,7}=4\frac{|V_{tb}V_{ts}^{\ast}|}{V_{tb}V_{ts}^{\ast}}\xi^{L,R}e^{-i\phi},
\end{eqnarray}
with
\begin{eqnarray}
\xi^{L,R}=\left(\frac{g^{\prime}M_Z}
 {g_1M_{Z^{\prime}}}\right)^2\left|\frac{B_{sb}^LB_{dd}^{L,R}}{V_{tb}V_{ts}^{\ast}}
 \right|.
\end{eqnarray}

To address the effect of $Z'$ boson, we have to know the values of
the $\Delta C_7$ and $\Delta C_9$ or equivalently $B_{sb}^L$ and
$B_{dd}^{L,R}$. Generally, we always expect $g'/g_1 \sim 1 $, if
both the $U(1)$ gauge groups have the same origin from some grand
unified theories. And $M_Z/M_{Z'} \sim 0.1 $ for  TeV scale neutral
$Z'$ boson, which yields $y \sim 10^{-2}$. In the first
paper of  Ref. \cite{Barger:2009hn} assuming a small mixing between
$Z-Z'$ bosons the value of $y$ is taken as $y \sim 10^{-3}$. In
order to explain the mass difference of $B_s - \bar B_s $ mixing, we
need $|B_{sb}^L| \sim |V_{tb} V_{ts}^*| $. Similarly,  the CP
asymmetry anomaly in $B \to \phi K, \pi K $ can be resolved if
$|B_{sb}^L B_{ss}^{L,R}| \sim |V_{tb} V_{ts}^*|$, which indicates
$|B_{ss}^L| \sim 1 $. Above issues have been discussed widely
in Ref. \cite{Cheung:2006tm}. Because we expect that $|B_{dd}^L| $
and $|B_{ss}^L| $ should have the same order of magnitude, we simply assume that
\begin{eqnarray}
|\xi|=|\xi_d^{R}|=|\xi_d^{L}|=\frac{1}{2}|\xi_u^{R}|
=\frac{1}{2}|\xi_u^{L}|\in(10^{-3},10^{-2}),
\end{eqnarray}
since the major objective of our work is searching for new physics signal,
rather than producing acute numerical results. Due to
renormalization group (RG) evolution from the $M_W$ scale to $m_b$
scale, the other Wilson coefficients also receive the contribution
of $Z^\prime$, however, the RG running from the $m_{Z^\prime}$ to $M_W$ scale has
been neglected in this work. The Wilson coefficients at $m_b$  and
$\sqrt{\Lambda_h m_b}$ scale have been presented in Table. 2.
\begin{table}[t]
 \begin{center}
 \caption{The Wilson coefficients $C_i$ within the SM and with the contribution
 from $Z^{\prime}$ boson included in NDR scheme at the scale $\mu=m_{b}$ and
 $\mu_h=\sqrt{\Lambda_h m_b}$.}
 \label{Wilson}
 \vspace{0.1cm}
 \small
 \doublerulesep 0.7pt \tabcolsep 0.04in
 \begin{tabular}{c|cc|cc}\hline\hline
 Wilson                   &\multicolumn{2}{c|}{$\mu=m_{b}$}       &\multicolumn{2}{c}{$\mu_h=\sqrt{\Lambda_h m_b}$} \\
 \cline{2-3}\cline{4-5}
 coefficients             &$C_i^{SM}$ &$\Delta C_i^{Z^{\prime}}$  &$C_i^{SM}$ &$\Delta C_i^{Z^{\prime}}$\\ \hline\hline
 $C_1$                    &$1.075$    &$-0.006\xi^L$      &$1.166$    &$-0.008\xi^L$\\
 $C_2$                    &$-0.170$   &$-0.009\xi^L$      &$-0.336$   &$-0.014\xi^L$ \\
 $C_3$                    &$0.013$    &$0.05\xi^L-0.01\xi^R$  &$0.025$    &$0.11\xi^L-0.02\xi^R$\\
 $C_4$                    &$-0.033$   &$-0.13\xi^L+0.01\xi^R$ &$-0.057$   &$-0.24\xi^L+0.02\xi^R$\\
 $C_5$                    &$0.008$    &$0.03\xi^L+0.01\xi^R$  &$0.011$    &$0.03\xi^L+0.02\xi^R$\\
 $C_6$                    &$-0.038$   &$-0.15\xi^L+0.01\xi^R$ &$-0.076$   &$-0.32\xi^L+0.04\xi^R$\\
 $C_7/{\alpha}_{em}$      &$-0.015$   &$4.18\xi^L-473\xi^R$   &$-0.034$   &$5.7\xi^L-459\xi^R$\\
 $C_8/{\alpha}_{em}$      &$0.045$    &$1.18\xi^L-166\xi^R $  &$0.089$    &$3.2\xi^L-355\xi^R$\\
 $C_9/{\alpha}_{em}$      &$-1.119$   &$-561\xi^L+4.52\xi^R$  &$-1.228$   &$-611\xi^L+6.7\xi^R$\\
 $C_{10}/{\alpha}_{em}$   &$0.190$    &$118\xi^L-0.5\xi^R$    &$0.356$    &$207\xi^L-1.4\xi^R$\\
  \hline \hline
 \end{tabular}
 \end{center}
 \end{table}

\begin{figure}
\begin{center}
\includegraphics[width=8cm,height=6cm]{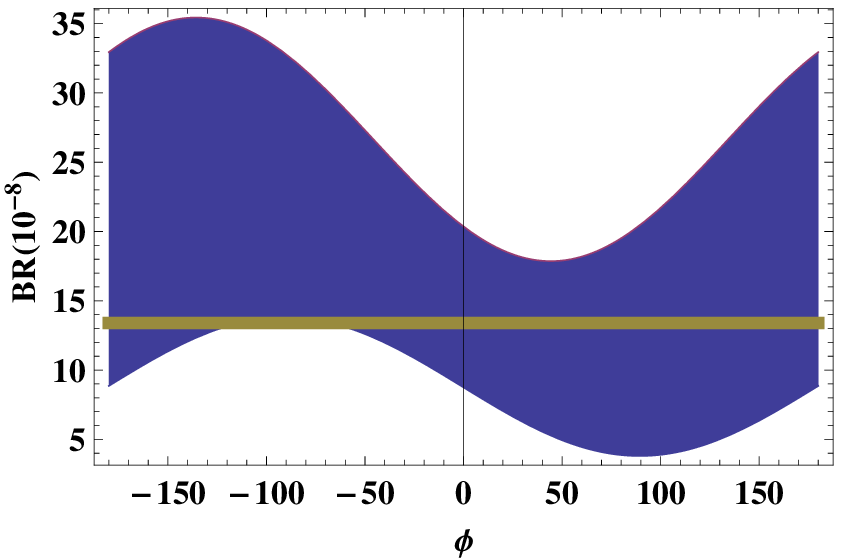}\,\,\,
\includegraphics[width=8cm,height=6cm]{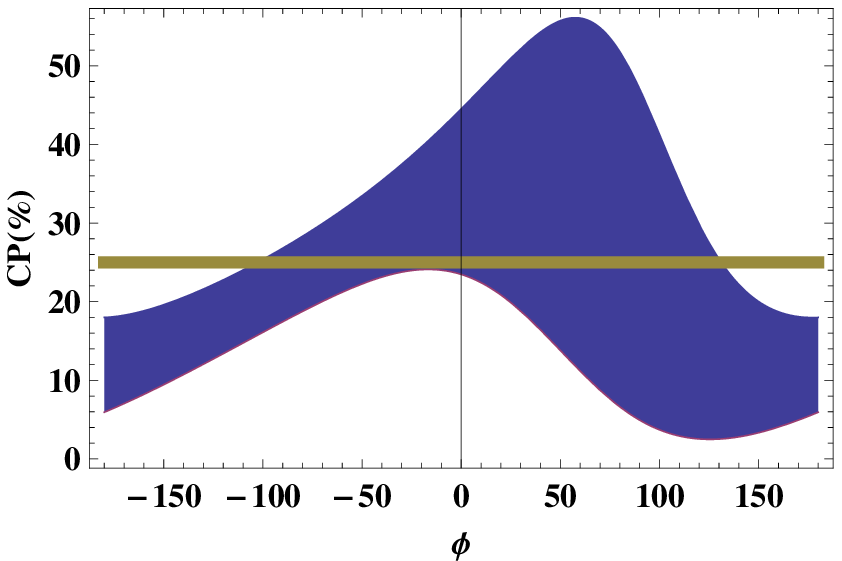}
\caption{After setting $\xi=0.01$, the variation of the CP averaged
branching ratio (left panel)  and  direct CP asymmetry
 (in \%) (right panel) as a function of  the new weak phase $\phi$.
 We varied the unitary angle
$\gamma \in (50^\circ,110^\circ )$. The horizontal lines are
predicted in the SM.} \label{x1}
\end{center}
\end{figure}

\begin{figure}
\begin{center}
\includegraphics[width=10cm,height=9cm]{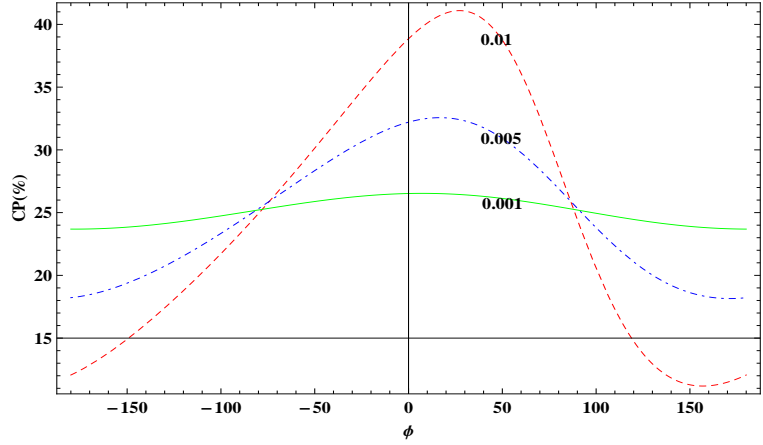}
\caption{When setting $\gamma=70^\circ$, the variation of  direct CP
asymmetry  with the new weak phase $\phi$, where the solid,
dot-dashed and dashed lines correspond to $\xi=0.001, 0.005$ and
$0.01$.} \label{x2}
\end{center}
\end{figure}

\begin{figure}
\begin{center}
\includegraphics[width=7.5cm,height=6cm]{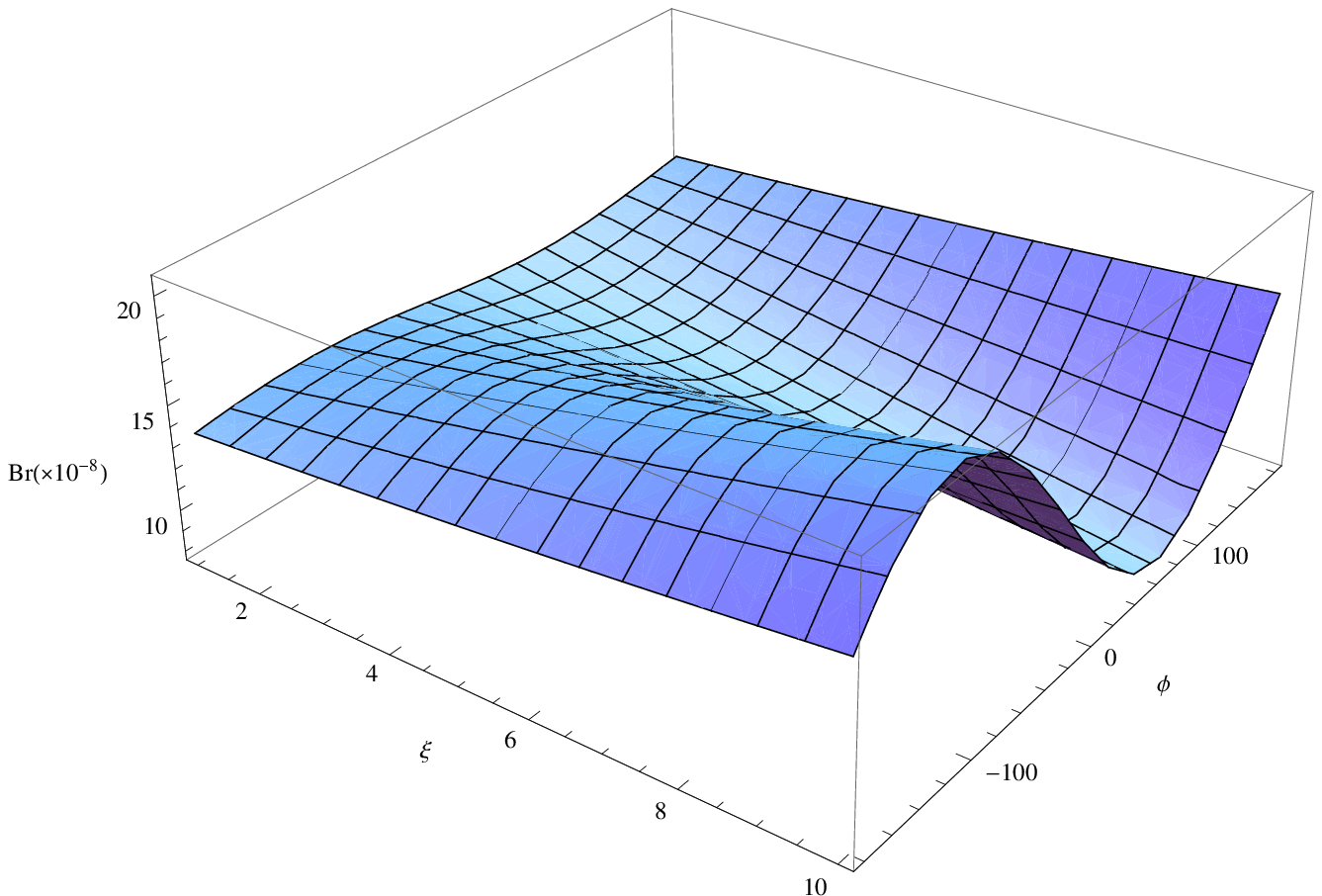}\,\,\,
\includegraphics[width=7.5cm,height=6cm]{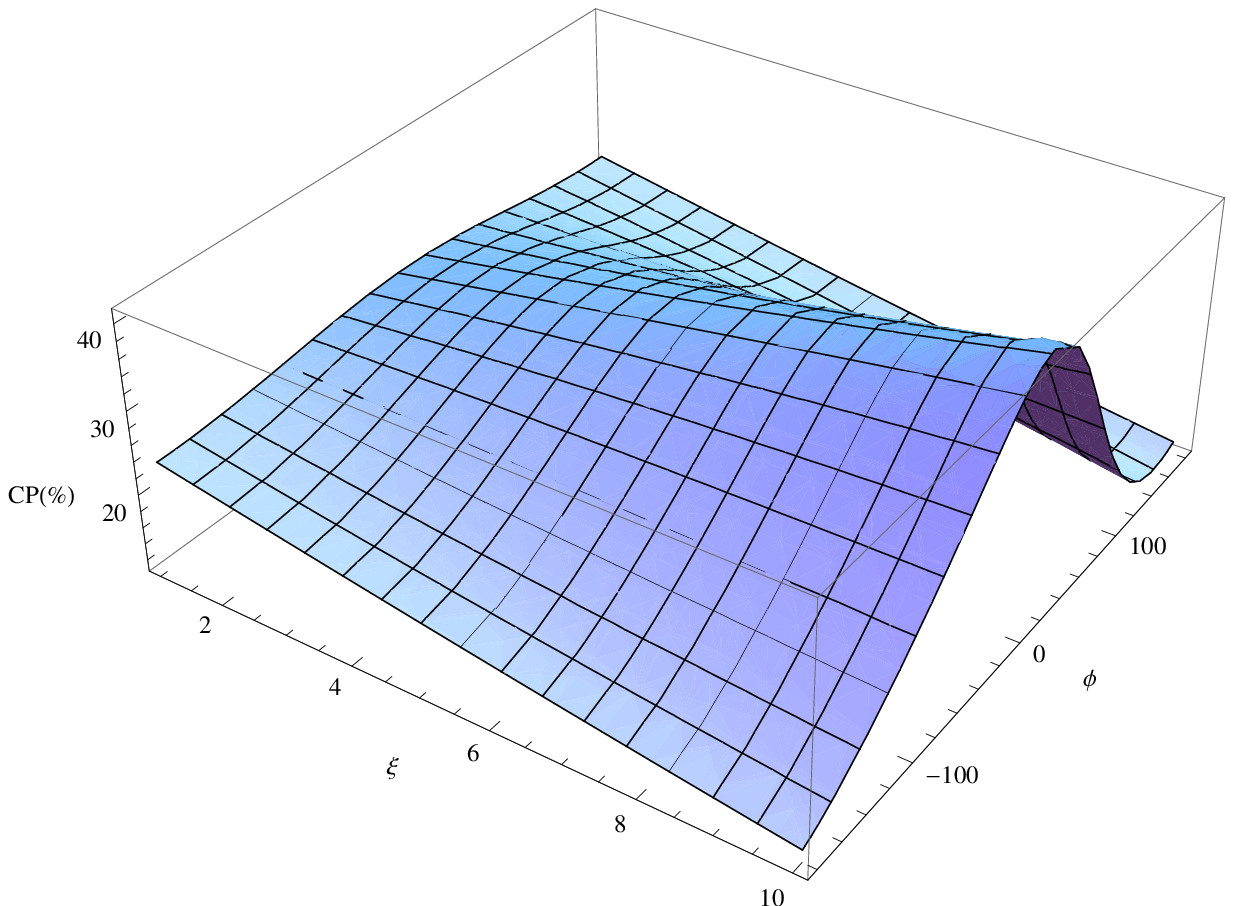}
\caption{The variation of the CP averaged branching ratio (left
panel) and the direct CP violation (right panel) with $\xi$ (in
units of $10^{-3}$) and the new weak phase $\phi$. } \label{x3}
\end{center}
\end{figure}


Once obtaining the values of the Wilson coefficients at the
scale $m_b$ and  $\sqrt{\Lambda_h m_b}$, we can get the decay
amplitude from the  $Z^\prime$, analogous to Eq. (\ref{eq:af1}), as:
\begin{multline} \label{eq:af3}
\Delta{\cal A}(\overline{B_s^0}(p_B)\to \phi(\epsilon,p_1)
\pi^0(p_2) )=- i\frac{G_F}{2} 2 m_\phi f_\pi(\epsilon^*\cdot p_B)
A_0^{B_s\to \phi}(0)\\ \times\Big[V_{ub}V_{us}^* (\Delta
a_2[u]+\Delta a_3[u]-\Delta a_3[d]-\Delta a_5[u]+\Delta
a_5[d]-\Delta a_7[u]
-\frac{1}{2}\Delta a_7[d]+\Delta a_9[u]+\frac{1}{2}\Delta a_9[d])\\
+V_{cb}V_{cs}^*(\Delta a_3[u]-\Delta a_3[d]-a_5[u]+\Delta
a_5[d]-\Delta a_7[u]-\frac{1}{2}\Delta a_7[d]+\Delta
a_9[u]+\frac{1}{2}\Delta a_9[d])\Big].
\end{multline}
To study the effect of the $Z^\prime$ boson, by setting $\xi=0.01$
and varying  $\gamma$ within $50^\circ$ to $110^\circ$, one can
get the variation of the CP averaged branching ratio and the direct
CP asymmetry as a function of the new weak phase $\phi$, as shown in
Fig. \ref{x1}, where the horizontal lines are the values predicted in the
SM. From these figures, we find that the branching ratio may become
three times of that predicted in the SM or drop to one third of the SM value within
the allowed parameter space. Moreover, as we mentioned before, we have
introduced one new weak phase $\phi$ from the off-diagonal element of $B_{sb}^{L}$,
which plays a major role in
changing the direct CP asymmetry. The direct CP violation can
reach $55\%$ if $\gamma =50^\circ$ and $\phi=70^\circ$. This
remarkable enhancement will be an important signal in testing the
model. Taking $\gamma=70^\circ$, we plot the variation of  direct CP
asymmetry as a function of the new weak phase $\phi$ with different
$\xi=0.001,0.005,0.01$, as shown in Fig. \ref{x2}. According to this
figure, we note that the new physics effect cannot be
detected if $\xi \leq 0.001$, namely a heavier $Z^\prime$ boson. If
there exists a light $Z^\prime$ boson, the observation of this mode
will in turn help us constraint the mass of $Z^\prime$. In
Fig. \ref{x3}, when leaving the $\xi$ and $\phi$ as free parameters,
and setting $\gamma=70^\circ$, we present the correlations between
the averaged branching ratio, direct CP asymmetry and the parameter
values by the three-dimensional scatter plots. As  illustrated in
Fig. \ref{x3}, the enhancement of both branching ratio and CP
asymmetry cannot be fulfilled at the same parameter values.

To conclude, we have calculated the branching ratio and direct CP
asymmetry of the decay mode $B_s \to \phi \pi^0$ within the QCD
factorization approach in both the SM and the non-universal $Z^\prime$ model. This
approach is suitable as the decay mode has no pollution from
annihilation diagrams. Upon calculation, we found the branching
ratio may be enlarged three times or decreased to one third  by the
effect of $Z^\prime$ boson within the allowed  parameter space.
Furthermore, as the direct CP asymmetry is concerned, it can reach
$55\%$ with a light $Z^\prime$ boson and suitable CKM phase. Also,
we note the enhancement of both branching ratio and CP asymmetry
cannot be accomplished at the same parameter space. Thus, if this mode
could be measured in the upcoming LHC-b experiment and/or Super B-factories it will provide a
signal of the non-universal $Z^\prime$ model, and can be used to constrain the mass of the
$Z^\prime$ boson in turn.

\section*{Acknowledgement}
The work of C.S.K. was supported in part by Basic Science Research
Program through the NRF of Korea funded by MOEST (2009-0088395) and
in part by KOSEF through the Joint Research Program (F01-2009-
000-10031-0). The work of Y.L. was supported by the Brain Korea 21
Project and by the National Science Foundation under contract
Nos.10805037 and 10625525.


\begin{thebibliography}{99}
\bibitem{Langacker:2000ju}
  P.~Langacker and M.~Plumacher,
  Phys.\ Rev.\  D {\bf 62}, 013006 (2000)
  [arXiv:hep-ph/0001204].

\bibitem{Buchalla:1995dp}
  G.~Buchalla, G.~Burdman, C.~T.~Hill and D.~Kominis,
  Phys.\ Rev.\  D {\bf 53}, 5185 (1996)
  [arXiv:hep-ph/9510376].

\bibitem{Nardi:1992nq}
  E.~Nardi,
  Phys.\ Rev.\  D {\bf 48}, 1240 (1993)
  [arXiv:hep-ph/9209223].

\bibitem{Barger:2009hn}
  V.~Barger, {\sl et. al},
  Phys.\ Lett.\  B {\bf 580}, 186 (2004)
  [arXiv:hep-ph/0310073];\\
  V.~Barger, {\sl et. al},
  Phys.\ Lett.\  B {\bf 598}, 218 (2004)
  [arXiv:hep-ph/0406126];\\
  V.~Barger, {\sl et. al},
  arXiv:0906.3745 [hep-ph];\\
 V.~Barger, {\sl et. al},
  Phys.\ Rev.\  D {\bf 80}, 055008 (2009)
  [arXiv:0902.4507 [hep-ph]].
\bibitem{Cheung:2006tm}
  K.~Cheung, {\sl et. al},
  Phys.\ Lett.\  B {\bf 652}, 285 (2007)
  [arXiv:hep-ph/0604223];\\
  C.~W.~Chiang, {\sl et. al},
  JHEP {\bf 0608}, 075 (2006)
  [arXiv:hep-ph/0606122];\\
  C.~H.~Chen and H.~Hatanaka,
  Phys.\ Rev.\  D {\bf 73}, 075003 (2006)
  [arXiv:hep-ph/0602140];\\
  Q.~Chang, X.~Q.~Li and Y.~D.~Yang,
  JHEP {\bf 0905}, 056 (2009)
  [arXiv:0903.0275 [hep-ph]];\\
  Q.~Chang, X.~Q.~Li and Y.~D.~Yang,
  arXiv:0907.4408 [hep-ph],\\
  C.~H.~Chen,
  arXiv:0911.3479 [hep-ph];\\
  C.~W.~Chiang, R.~H.~Li and C.~D.~Lu,
  arXiv:0911.2399 [hep-ph];\\
  R.~Mohanta and A.~K.~Giri,
  Phys.\ Rev.\  D {\bf 79}, 057902 (2009)
  [arXiv:0812.1842 [hep-ph]].

\bibitem{Langacker:2008yv}
  P.~Langacker,
  arXiv:0801.1345 [hep-ph].

\bibitem{Beneke:1999br}
  M.~Beneke, G.~Buchalla, M.~Neubert and C.~T.~Sachrajda,
  Phys.\ Rev.\ Lett.\  {\bf 83}, 1914 (1999)
  [arXiv:hep-ph/9905312];\\
 M.~Beneke, G.~Buchalla, M.~Neubert and C.~T.~Sachrajda,
  Nucl.\ Phys.\  B {\bf 591}, 313 (2000)
  [arXiv:hep-ph/0006124].

\bibitem{Beneke:2003zv}
  M.~Beneke and M.~Neubert,
  Nucl.\ Phys.\  B {\bf 675}, 333 (2003)
  [arXiv:hep-ph/0308039].


\bibitem{PQCD}
  A.~Ali, {\sl et.al},
  Phys.\ Rev.\  D {\bf 76}, 074018 (2007)
  [arXiv:hep-ph/0703162].

\bibitem{Buras} For a review, see G. Buchalla, A.J. Buras,
M.E. Lautenbacher, Rev. Mod. Phys. {\bf68}, 1125 (1996).

\bibitem{chaudhuri}
  S.~Chaudhuri, S.~W.~Chung, G.~Hockney and J.~Lykken,
  Nucl.\ Phys.\ B {\bf 456}, 89 (1995);\\
  G.~Cleaver, M.~Cvetic, J.~R.~Espinosa, L.~L.~Everett, P.~Langacker and
  J.~Wang,
  Phys.\ Rev.\ D {\bf 59}, 055005 (1999);\\
  M.~Cvetic, G.~Shiu and A.~M.~Uranga,
  Phys.\ Rev.\ Lett.\  {\bf 87}, 201801 (2001);\\
  M.~Cvetic, P.~Langacker and G.~Shiu,
  Phys.\ Rev.\ D {\bf 66}, 066004 (2002).

\bibitem{gut}
T.K.~Kuo, N.~Nkagawam, Phys.\ Rev.\  D {\bf 66}, 066004 (1984);\\
V.D.~Barger, {\sl et.al}, Int.\ J.\ Mod.\  A {\bf 2}, 1327 (1987).
\bibitem{ajburas}
A.J.~Buras, R.~Fleischer, S.~Recksiegel and F.~Schwab, Phys. Rev.
Lett. {\bf 92}, 101804 (2004)
\end{thebibliography}
\end{document}